\documentclass[Royal,sageh,times]{sagej}

\usepackage{moreverb,url}

\usepackage[colorlinks,bookmarksopen,bookmarksnumbered,citecolor=red,urlcolor=red]{hyperref}

\def\volumeyear{2018}

\begin{document}

\runninghead{Randon-Furling, Olteanu, Lucquiaud}

\title{From urban segregation to spatial structure detection}

\author{Julien Randon-Furling\affilnum{1}, Madalina Olteanu\affilnum{1,2}, Antoine Lucquiaud\affilnum{1}}

\affiliation{\affilnum{1} SAMM (EA4543), Universit\'e Paris 1 Panth\'eon Sorbonne\\
\affilnum{2} MaIAGE, INRA, Universit\'e Paris-Saclay}

\corrauth{Julien Randon-Furling, SAMM (EA4543)
Universit\'e Paris 1 Panth\'eon Sorbonne,
Centre Pierre Mend\`es-France,
90 rue de Tolbiac,
75013 Paris, FRANCE
}

\email{Julien.Randon-Furling@univ-paris1.fr}

\begin{abstract}
We develop a ``multifocal'' approach to reveal spatial dissimilarities in cities, from the most local scale to the metropolitan one. Think for instance of a statistical variable that may be measured at different scales, \textit{eg} ethnic group proportions, social housing rate, income distribution, or public transportation network density. Then, to any point in the city there corresponds a sequence of values for the variable, as one zooms out around the starting point, all the way up to the whole city --~as if with a varifocal camera lens. The sequences thus produced encode in a precise manner spatial dissimilarities: how much they differ from perfectly random sequences is indeed a signature of the underlying spatial structure. We introduce here a mathematical framework that allows to analyze this signature and we provide a number of illustrative examples.
\end{abstract}


\maketitle

\section{Introduction}
Cities may be regarded as an epitome of complex systems (\cite{batty07}): individual and institutional agents interact on multiple levels of numerous networks (both physical and virtual), leading to nontrivial collective behaviour and nontrivial patterns at many scales. Among a number of intricate questions that have emerged in the study of urban systems, that of sociospatial dissimilarities stands out as one that has aroused interest for decades --~ and across a wide range of fields thanks to, especially, the pioneering multi-agent models of segregation developed by~\cite{Schel1} and \cite{Sak2}. Urban segregation phenomena will serve here as archetypal examples to illustrate the mathematical methods and tools we are presenting.

Segregation is often simply perceived as spatial separation of two or more groups, and therefore measured in terms of the relative proportions of each group in the different neighbourhoods of a city. However, segregation is essentially a spatial and multiscalar phenomenon, as pointed out for instance by~\cite{Leckie12,Osth15,BarthLouf}. An individual perceives segregation all the more acutely as she has to go \textit{a longer way} from her home to discover what the city in its entirety might look like. Imagine the extreme case of a city where two groups $A$ and $B$ live in total separation, thus forming two ghettos. An individual living at the heart of one of the ghettos would have to explore the whole city to find out that it actually comprises equal proportions of both groups. On the contrary, starting from some parts of the city that are \textit{better mixed}, for instance on the boundary between the two ghettos, there would be no need to cover so large an area to come to the same realization. This basic observation has led us to imagine a mathematical framework that allows to capture and measure spatial dissimilarities as a multiscalar phenomenon across the city.

Consider ever larger neighbourhoods around a starting point or areal unit. As detailed in the first Section below, enlarging the area all the way up to the whole city produces for each point in the city a trajectory. That is, the sequence of values taken by the variable under consideration from its value at the most local level around the starting point to its average value at the metropolitan level. For a given starting point, how ``long'' it takes for the sequence to reach the city's average thus indicates how ``distant'' the point is from the city's global distribution.\\
Further, the same mapping of points into trajectories allows to easily characterize a null model. In a perfectly mixed city (the ``uniform'' city), the aggregation process used to build the trajectories would be equivalent to adding units drawn at random (as when drawing balls in an urn without replacement). We introduce in the second Section below various ways of quantifying deviations from such random sequences. In both sections, we illustrate our ideas and methods with public data available for the city of Paris.

\section{From bespoke neighbourhoods to trajectories}

Data is often available at a given spatial level --~for example census blocks in the USA or IRIS (\textit{\^Ilots regroup\'es pour l'information statistique}) in France, see ~\cite{census94} and~\cite{iris}. Aggregating such blocks poses a number of statistical questions, the most famous of which is maybe the so-called ``modifiable areal unit problem'', as stated in~\cite{OpT,Op}. Also, in most instances, elementary spatial units will exhibit dissimilarities. Indeed, if the data gives the number of medical practitioners, or the quantiles of the income distribution within the unit, then units will generally differ from one another~\cite{Glaes}. Such dissimilarities may or may not present spatial patterns. If they do, the geographical system may be said to exhibit spatial segregation, especially if the variables considered correspond to the relative proportions of different population groups:~\cite{cowgill51,duncan55,farley68,cortese76,schwartz80,white83,james85,masden88,reardon02,brown06,feitosa07,feitosa12,fossett17,caridi17,OTRF17}.

\begin{figure}
\begin{minipage}[c]{.22\linewidth}
\begin{center}
\includegraphics[height=2.1cm]{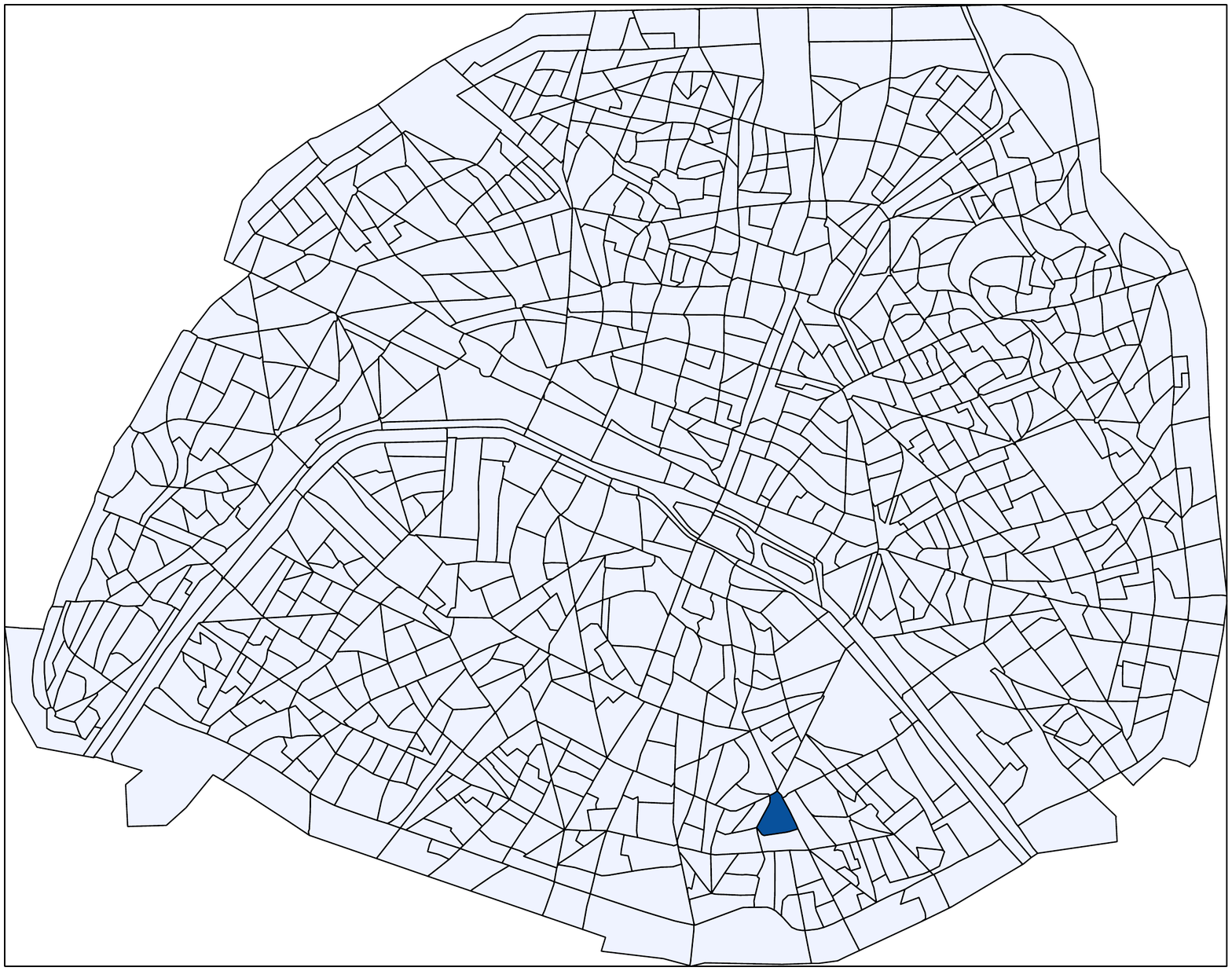}
\smallskip
\includegraphics[height=2.1cm]{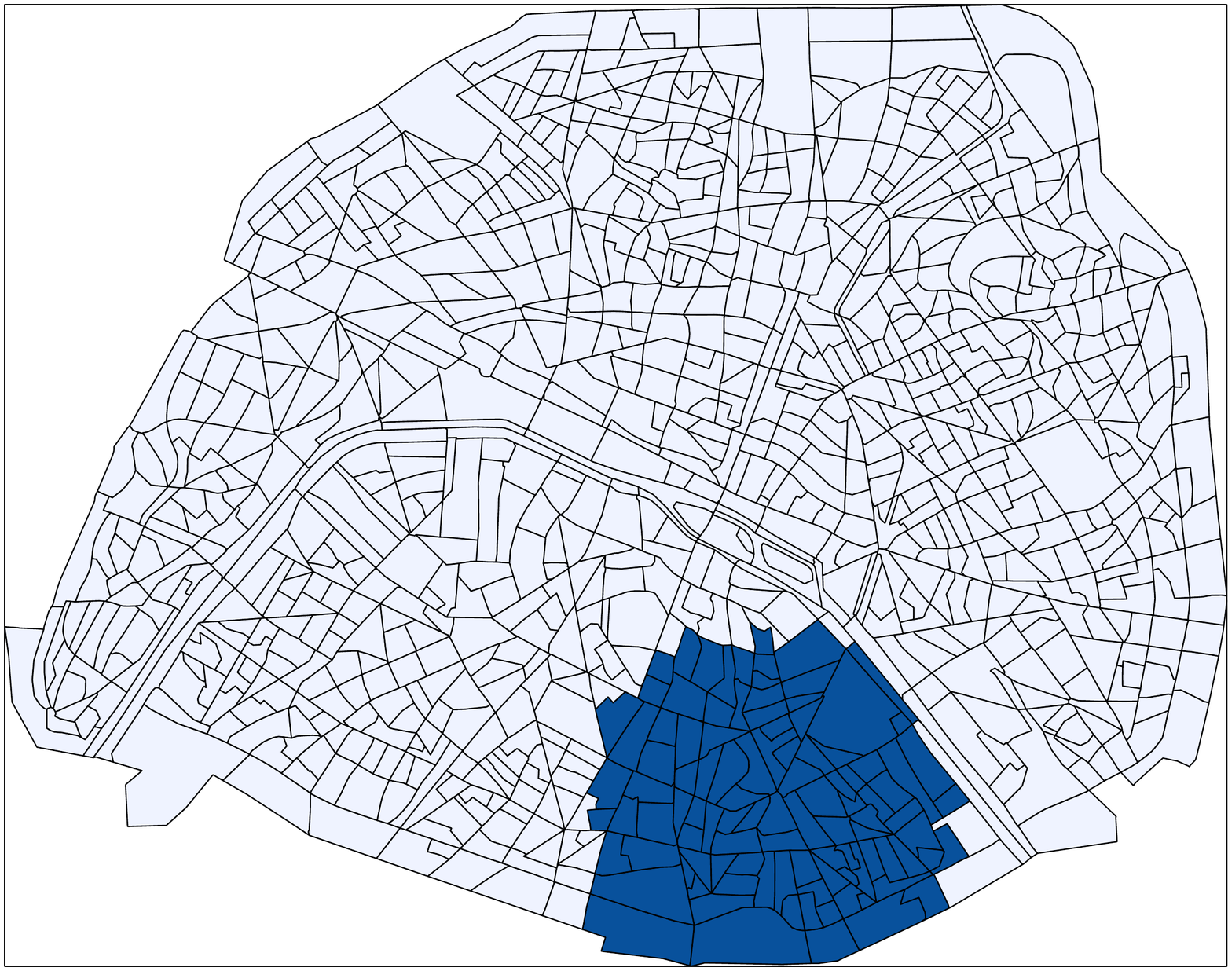}
\end{center}
   \end{minipage}
\begin{minipage}[c]{.22\linewidth}
\begin{center}
\includegraphics[height=2.1cm]{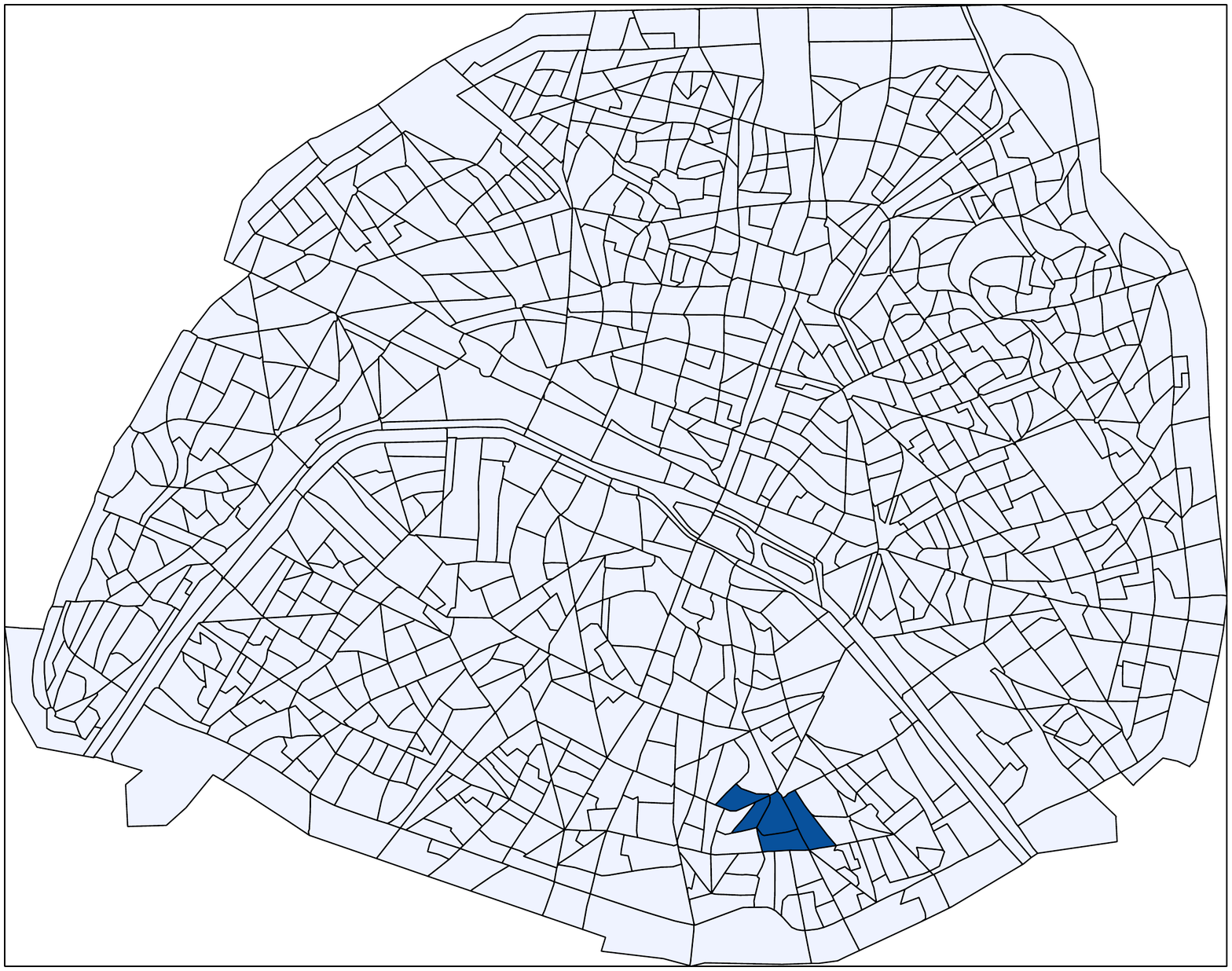}
\smallskip
\includegraphics[height=2.1cm]{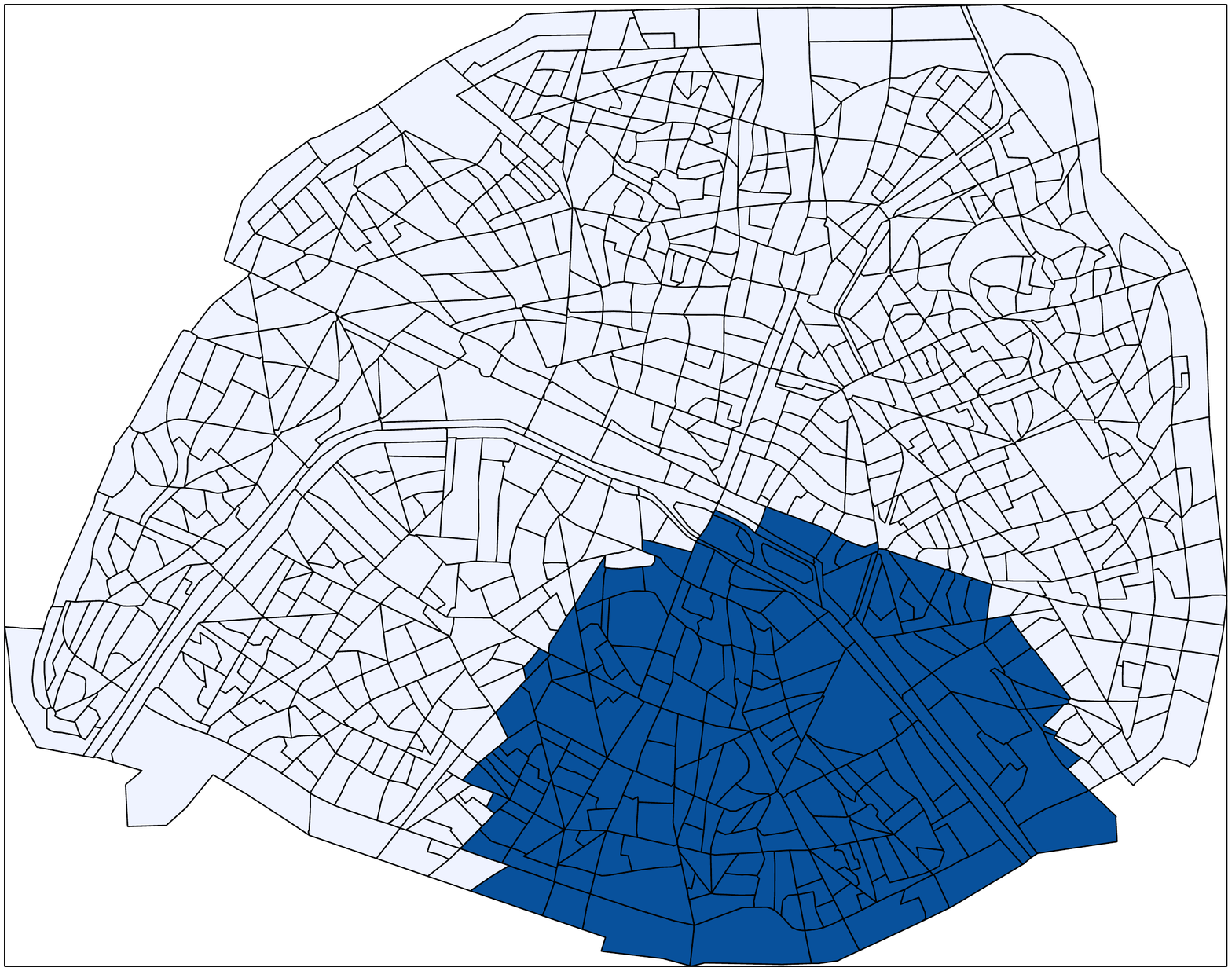}
\end{center}
   \end{minipage}
    \begin{minipage}[c]{.22\linewidth}
\begin{center}
\includegraphics[height=2.1cm]{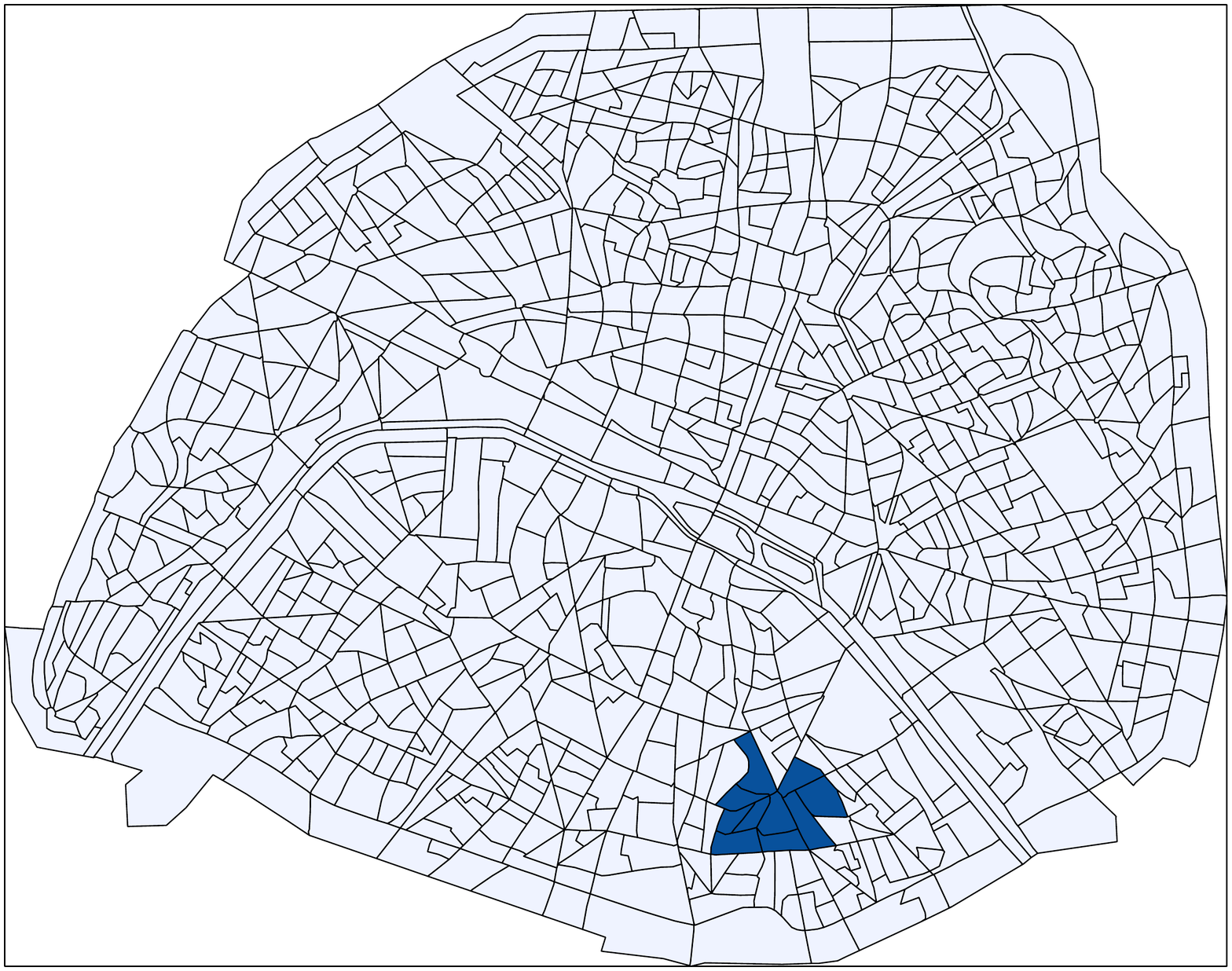}
\smallskip
\includegraphics[height=2.1cm]{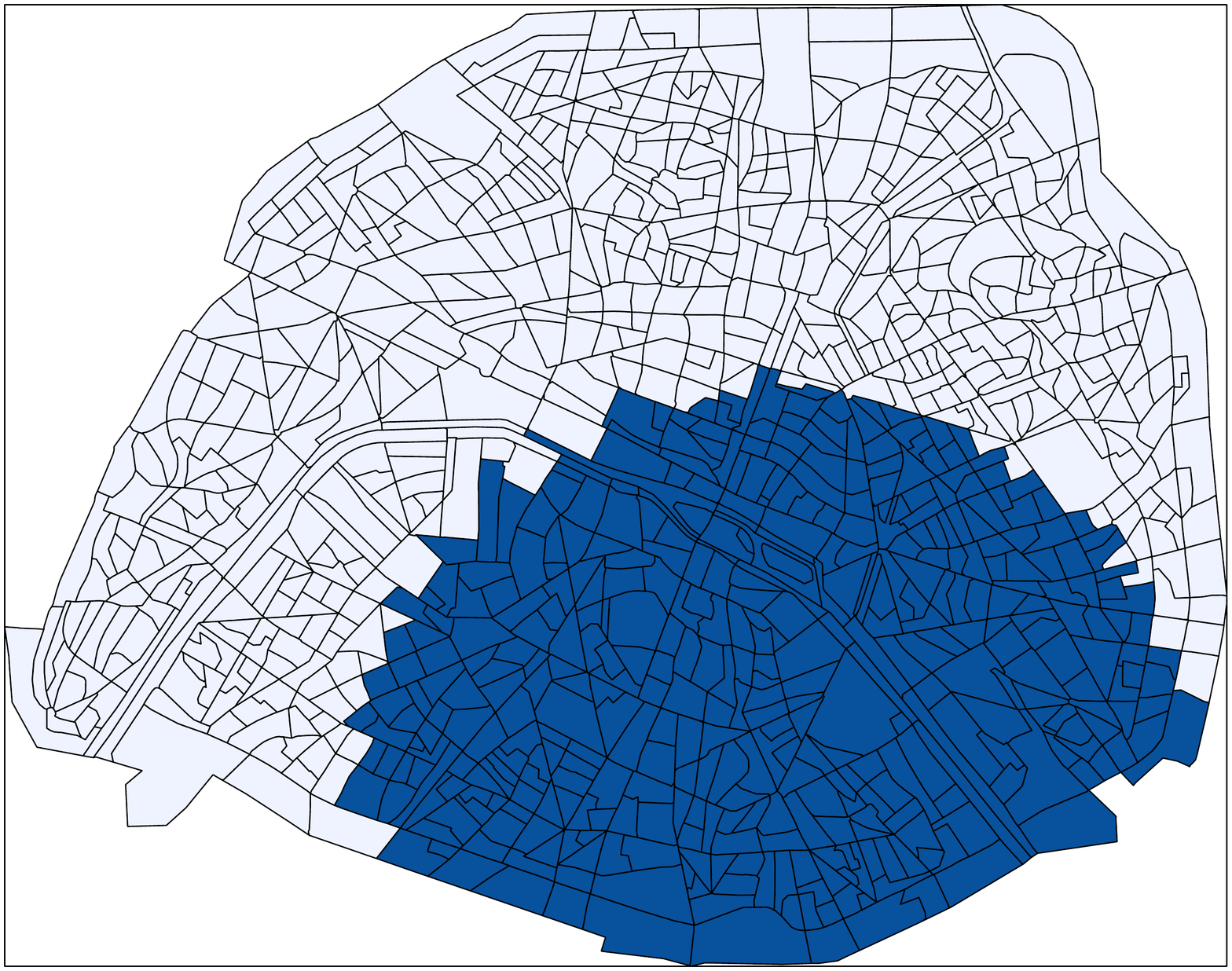}
\end{center}
   \end{minipage}
\begin{minipage}[c]{.22\linewidth}
\begin{center}
\includegraphics[height=2.1cm]{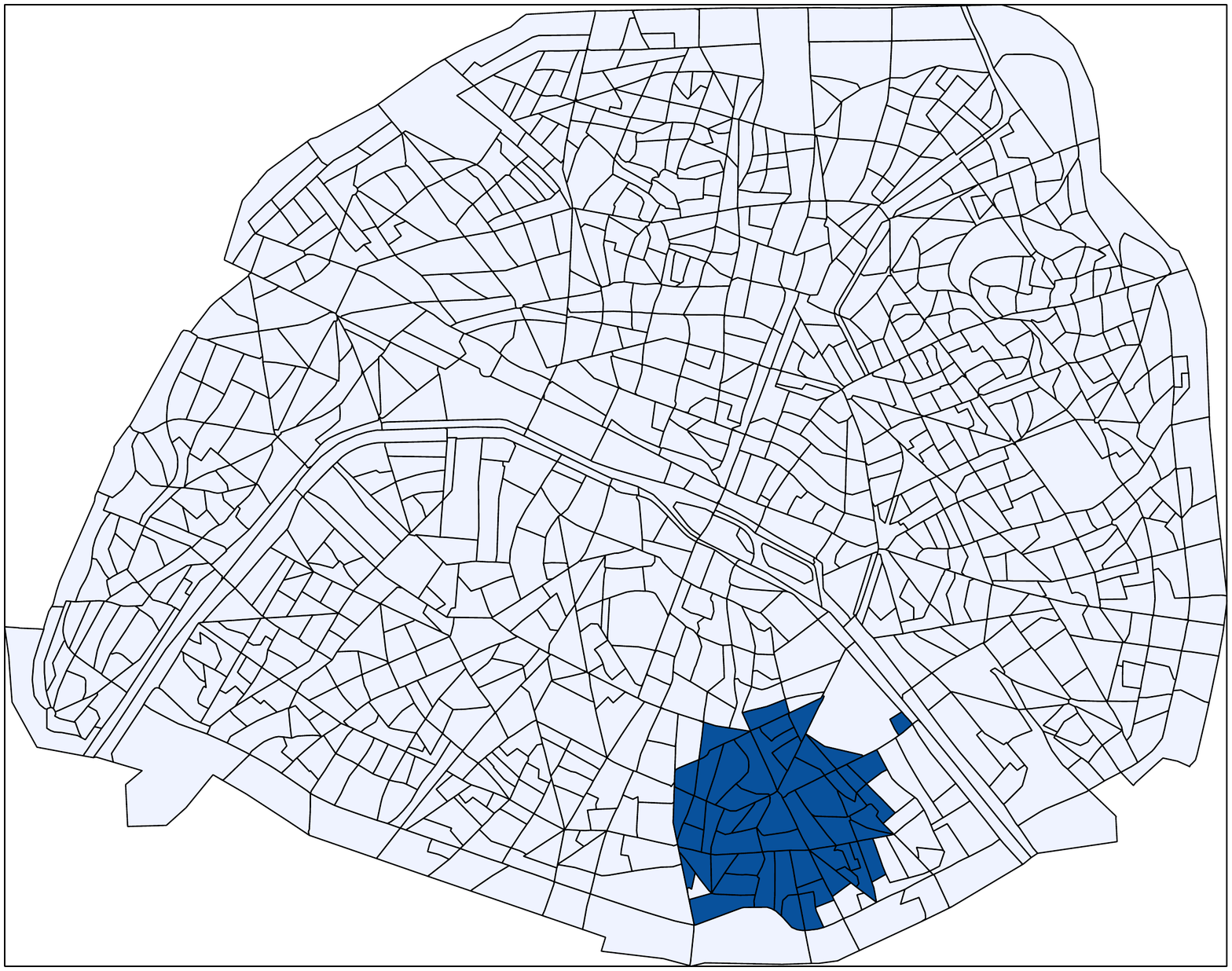}
\smallskip
\includegraphics[height=2.1cm]{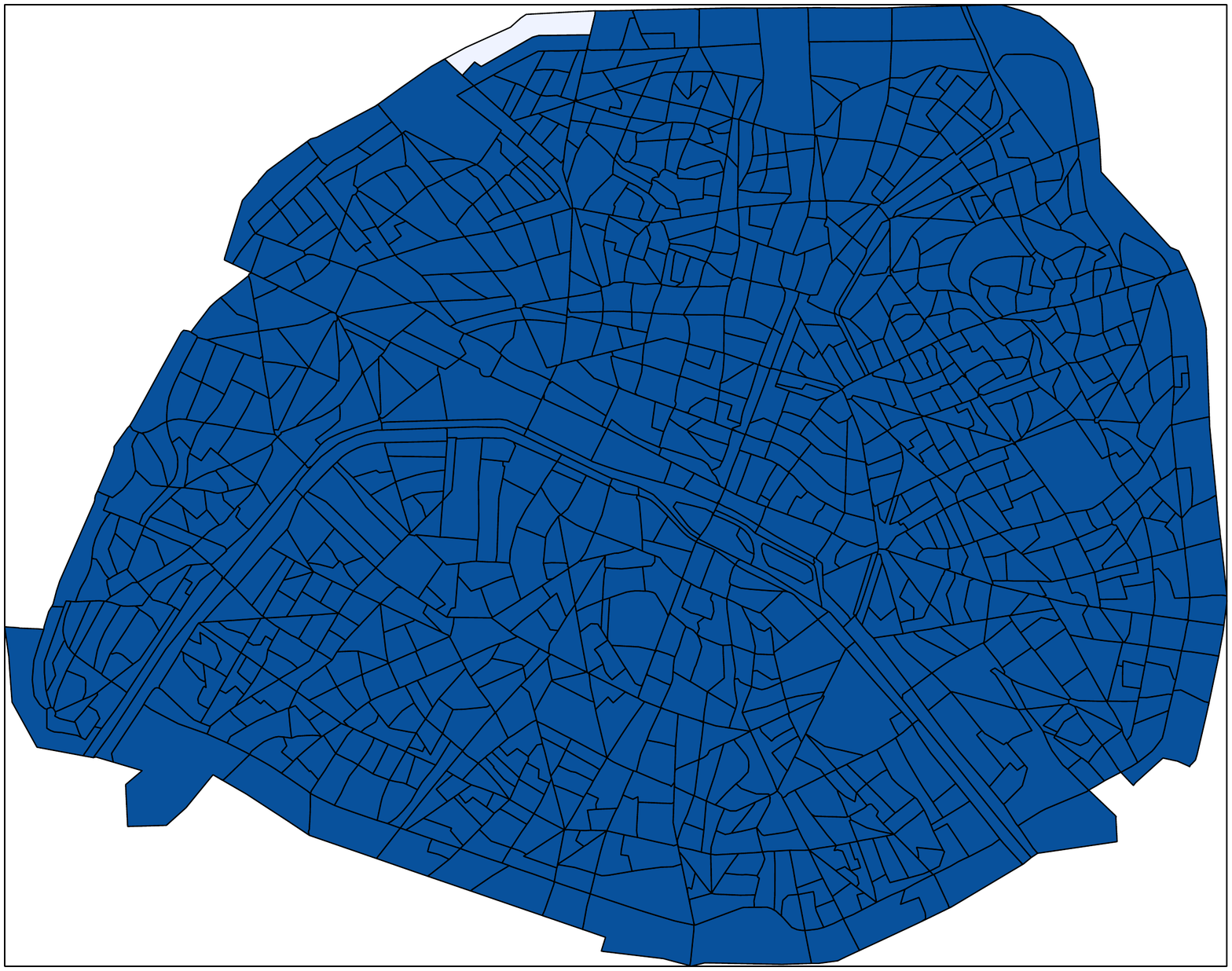}
\end{center}
   \end{minipage}
\centering
\caption{Some of the steps in a sequence of bespoke neighbourhoods centered around one of Paris's statistical units (so-called IRIS). The first point in the sequence corresponds to the unit alone (top left). A second unit is then aggregated to the first one, according to some prescribed rule --~here for the sake of simplicity, we choose the nearest neighbouring unit in terms of distance from centroid to centroid. The procedure is iterated until eventually one has aggregated all the city's units around the first one. This yields, for each starting point, a sequence of bespoke neighbourhoods, representing all scales, from the most local one available in the data to the metropolitan one.}
\label{fig:seq}
\end{figure}

Modifiable areal units may actually prove to be a valuable tool to quantify segregation and spatial patterns of dissimilarities, see~\cite{Leckie12,Clark15,Jones15,Manley15,Leckie15,Harris17,Adr17}. Instead of only comparing individual units, one may gain greater insight into relative spatial differences by taking into account the broader picture: that is, by considering ever larger neighbourhoods around a unit, as if through a zoom lens --~see~Figure~\ref{fig:seq}. Evaluating a given statistical variable at each step of the aggregation procedure produces a sequence of values that reflects not only the initial point but also the singularity of its position within the city. The sequence of bespoke neighbourhoods around a starting point acts as a sequence of filters: unveiling step by step the city's ``face''. Indeed the final value of all trajectories, whatever the starting point, will be the city's average for the variable considered. But details of a given trajectory, such as the number of aggregating steps needed to converge to the  city's average characterize how different, how secluded from the whole city the starting point is.

Although relatively easy to describe, the actual computing of trajectories presents a number of difficulties from a statistical point of view. We recall some of them in the following paragraphs.

\subsection{Defining and computing trajectories\label{sec:ETS}}

Consider a grid comprising $N$ fundamental spatial units (typically, individuals or the smallest statistical units available). For a given unit, define~$G_n(i)$, the cluster formed by itself and its $n-1$ nearest neighbouring units. Thus, $G_1(i)$ is just the $i$-th unit on its own, while $G_N(i)$ is the whole grid (whichever $i$ from which one starts).

Now suppose one has a statistical variable~$\xi$ defined on each element of $$\mathcal{G}=\left\lbrace G_n(i),\, 1\leq i,n \leq N\right\rbrace$$
Note that $\mathcal{G}$ is simply the set of all bespoke neighbourhoods, with all possible starting points in the city. Then, define for each starting point~$i$ a function~$f_i$ such that $f_i(n)$ gives the value of $\xi$ computed on~$G_n(i)$.

Interpreting $n$ as an index, one has for each starting unit~$i$ a sequence representing the trajectory that takes $\xi$ from its value on unit~$i$ to its value on the whole grid. Formally, $\left(f_i(n)\right)_{1\leq n \leq N}$ is what we call the trajectory of $i$ for the variable $\xi$.\\
Once trajectories have been built from the data available in a given city, a statistical analysis may be carried out in order to:
\begin{itemize}
\item identify units that exhibit similar trajectories --~this may be done for instance using clustering algorithms;
\item detect the crossing over between two regimes: the local one and the global one --~in the former, $\xi$ may take values significantly distinct from the one on the whole grid; in the latter, it takes values very similar to the city's average. 
\end{itemize}

Let us emphasize links between the first point above and multiscalar approaches used in recent papers, \textit{eg}~\cite{Osth15,Clark15,And15}. In particular, \cite{And15} use vertical slices or cross-sections of the trajectories defined here, i.e. values at certain points only. But looking at full trajectories allows us to examine the second point listed above: the number of aggregated neighbouring blocks needed around a starting block to get close enough to the city's average value for the variable under consideration. We shall call this the ``\textit{radius of convergence}''. Geographically this plays the role of a local urban radius: it is a proxy for the area one needs to explore locally to obtain a reasonably good perception of the city as a whole.
We give examples of such analysis further in this section.

\subsection{Constructing trajectories from actual data}
\label{sec:build}
\begin{figure}
\centering
\includegraphics[height=7cm]{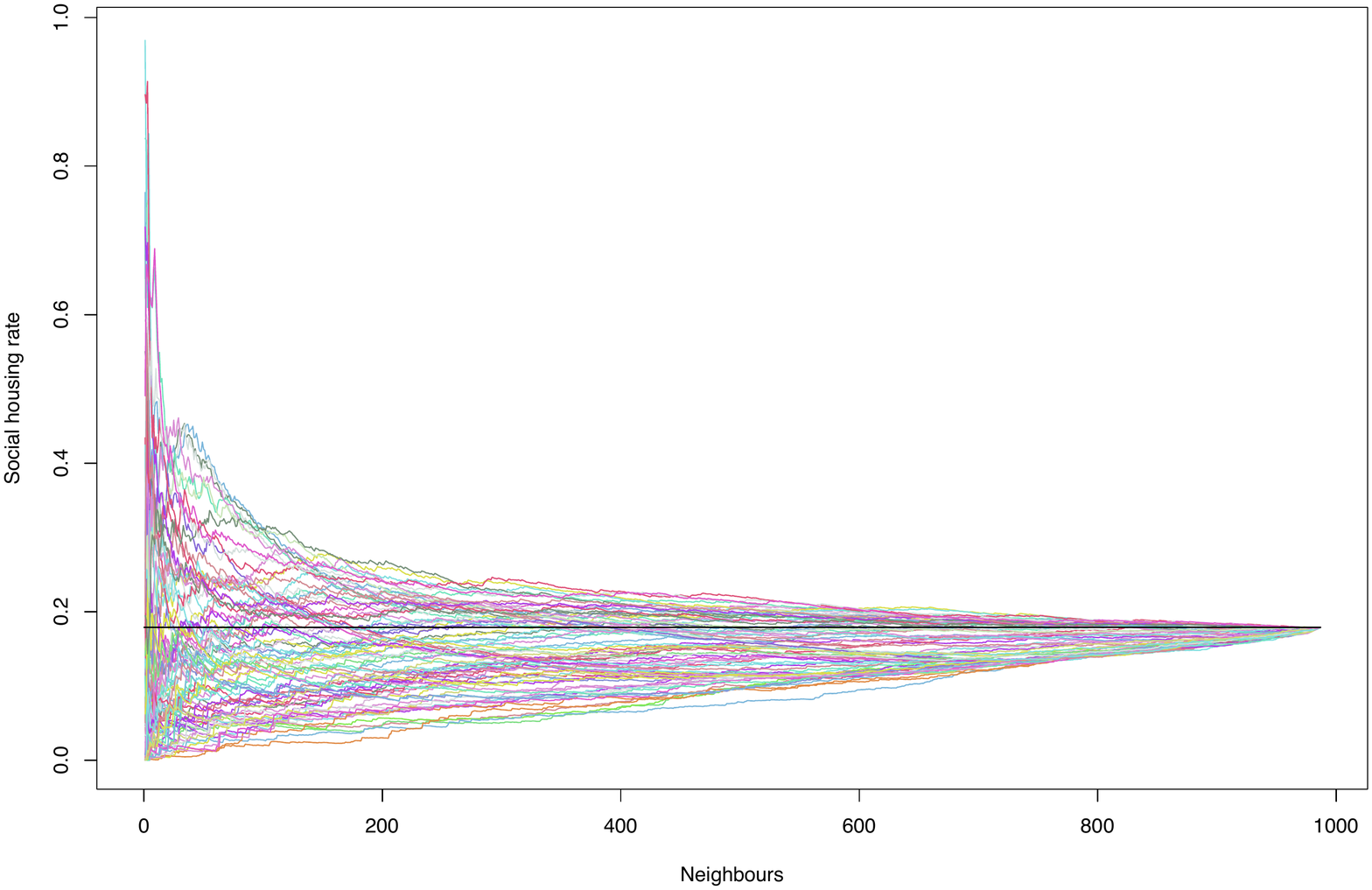}
\caption{Trajectories for the social housing rate, starting from some ($10$\%) of the 936 statistical units (so-called IRIS) in Paris. The solid line corresponds to the city's average ($17.9$\%).}
\label{fig:example}
\end{figure}

One of the statistical subtleties arises upon building datasets so that they are defined on $\mathcal{G}$, the full set of all possible bespoke neighbourhoods in the city.

Consider indeed the following standard situation: data is available in the form of a size factor $s_i$ (\textit{eg} a number of inhabitants) for each spatial unit, as well as the value of a possibly multi- or even infinite dimensional variable~$\xi$. This could be the number of offices or services of such and such type available in the unit, the social housing rate, an average income, a local density of public transportation, quantiles of a distribution, a full distribution...

When grouping $n$ units, one needs to compute the value of the variable on the new aggregate. This is done easily in the case when $\xi$ is simply a number:
\begin{equation}
\xi(G_n(i))=\sum_{j\in G_n(i)}\, \xi(j).
\end{equation}

Similarly, $\xi(G_n(i))$ is readily computed when $\xi(i)$ is a rate or an average:
\begin{equation}
\xi(G_n(i))=\frac{\sum_{j\in G_n(i)}\, s_j \xi(j)}{\sum_{j\in G_n(i)}\, s_j} \; ,
\end{equation}
with $s_i$ the population of unit $i$ (or another relevant size factor).

When $\xi$ is a distribution that is known entirely (\textit{eg} one knows the income of every single household in any of the units), one simply aggregates the individual datasets to obtain the dataset for a group of spatial units, and the corresponding empirical distribution is obtained readily.

A more difficult case, alas very frequent, is that when $\xi$ retains only certain percentiles of a distribution. For instance:
\begin{equation}
\xi(i)= \left\lbrace 2354; 4684; 6546; 8138; 10542; 13622; 17058; 22202; 30862\right\rbrace,
\end{equation}
are the deciles of the income distribution (in euros per year) for a given IRIS (census block) in the northern part of Paris.

In this case, one is led to either (or both) rely on an Ansatz for the shape of the underlying distribution (\textit{eg }assume that it is log-normal, or exponential or other) or simulate the full dataset (with assumptions on the intradecile distributions), see~\cite{stoker84,genest86,fother91,genest92,jelinski96,busetti14}. An example is given in the next Section, as we are faced with this problem when we consider income data for the city of Paris, available only in the form of quantiles for each census block.

Lastly, one does not have to be working in the discrete framework of statistical blocks or population count data. For certain variables, such as availability of public transport networks, one may work directly in terms of a local density. For instance, let $M$ be the total number of metro stations in Paris. Let $m_i(l)$ be the number of stations in a disk of radius $l$ centered on IRIS $i$, $p_i(l)$ the population living in the same disk, and $P$ the total population in Paris. Then define
\begin{equation}
r_i(l)=\frac{m_i(l)/M}{p_i(l)/P}=\frac{m_i(l)/p_i(l)}{M/P}.
\end{equation}
This is similar to the ``representation'' defined for social classes in~\cite{BarthLouf}. It quantifies whether the local density of metro stations per inhabitant is smaller ($r_i(l)<1$), larger ($r_i(l)>1$) or equal ($r_i(l)=1$) to the city's average density. And it allows to build trajectories, for each IRIS, letting $l$ vary from some small value $\varepsilon$ to the full radius of the city.

\subsection{Example: spatial dissimilarities in Paris\label{sec:PSH}}

As an illustration, let us work with two types of data available for the city of Paris from France's census bureau, called INSEE (``\textit{Institut national de la statistique et des \'etudes \'economiques}''). 

\subsubsection{Social housing rate.}

For each IRIS, the number of housing units and the number, among these, of social housing ones are available. One is then in a situation where computing the social housing rate for any group of IRIS is easy, and the trajectories can be computed for all IRIS in Paris. Some of these trajectories are shown in Figure~\ref{fig:example}.

One observes groups of trajectories that tend to start higher or lower than the metropolitan average, and converge to it more or less quickly, obviously with a strong spatial dependency as far as the initial units are concerned.

\begin{figure}
\centering
\begin{minipage}[c]{.45\linewidth}
\begin{center}
\includegraphics[height=4cm]{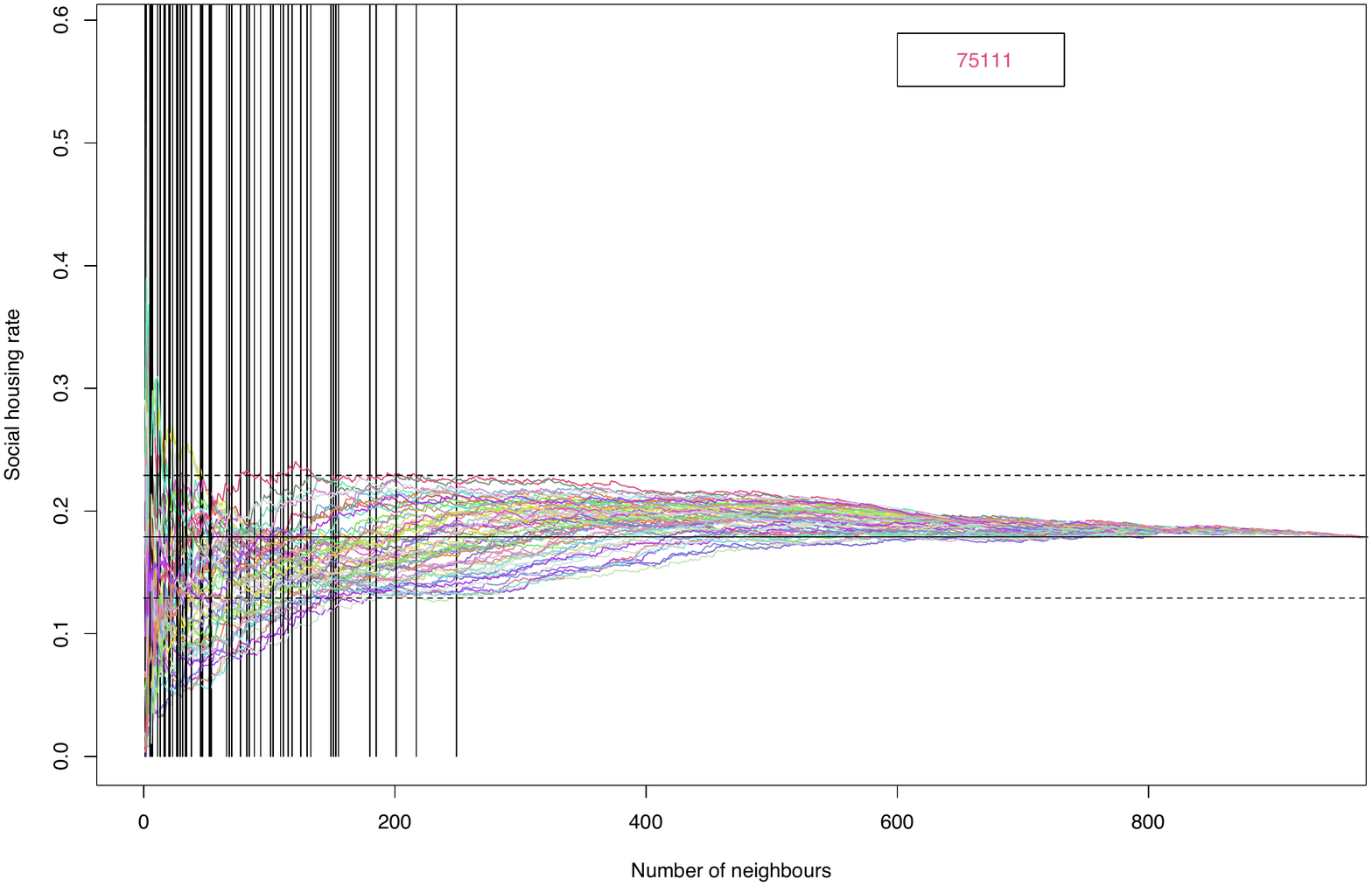}
\end{center}
   \end{minipage}
\begin{minipage}[c]{.45\linewidth}
\begin{center}
\includegraphics[height=4cm]{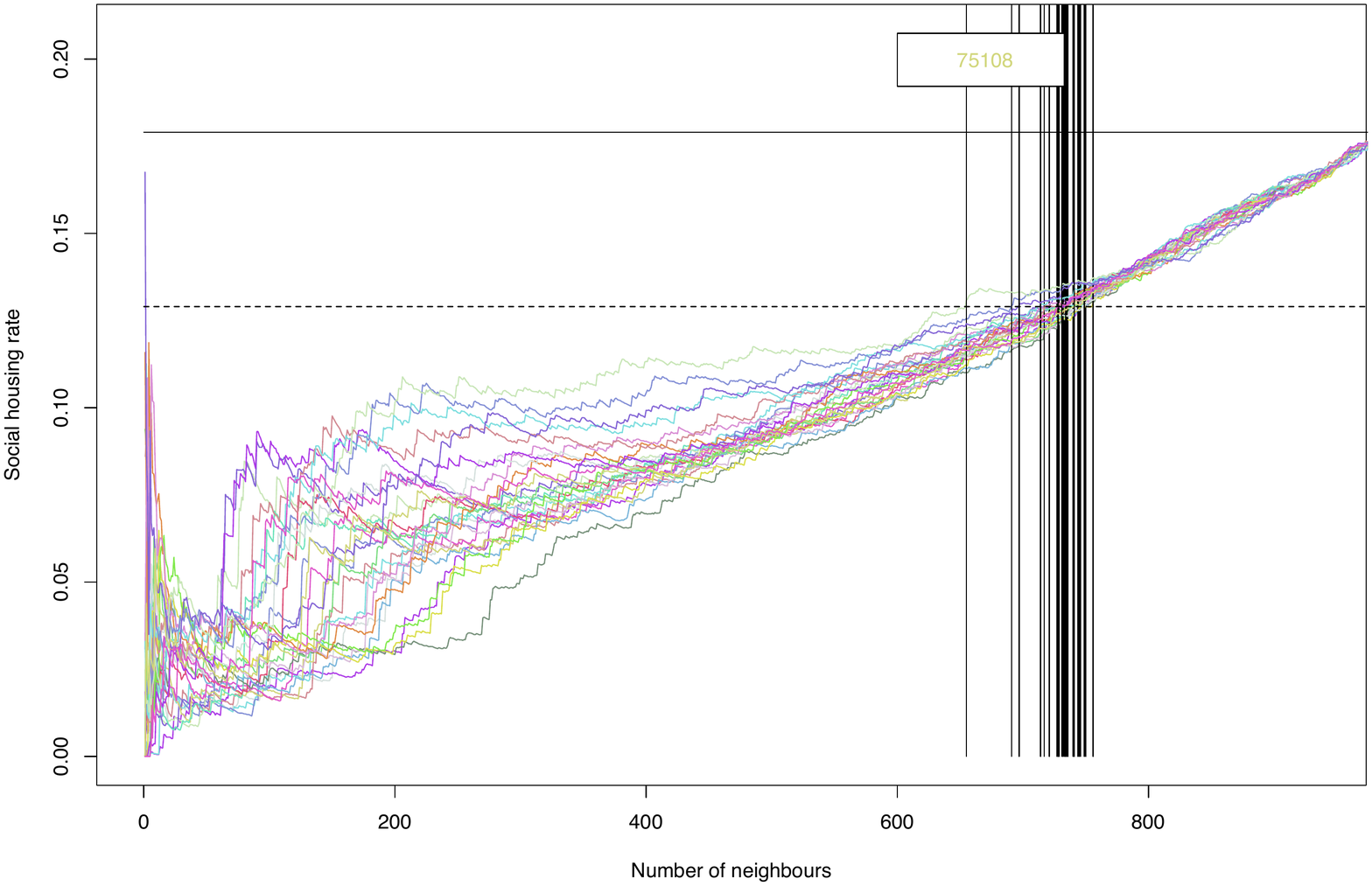}
\end{center}
   \end{minipage}
\caption{Trajectories for the social housing rate, starting from each IRIS (census block) in Paris's 11th (left) and 8th (right) districts. Colours correspond to each different IRIS taken as starting point. The solid flat line gives the city's average social housing rate ($17.9$\%), the dashed lines correspond to $\pm0.05$ around this average. Solid vertical lines correspond to radii of convergence: the radius of convergence of a trajectory is defined as the point where the path last enters the $\pm0.05$-interval (and therefore remains inside the interval afterwards). One observes that IRIS in the 11th district converge much faster to the city's average than IRIS in the 8th district.}
\label{fig:conv1}
\end{figure}

Let us define the radius of convergence, that is, the point where each path enters (and remains) into a given interval (here $\pm 0.05$) around the city's average social housing rate. Looking at these radii reveals different scales of convergence to the global mean, from one district to another and, inside each district, from one IRIS to another, as can be seen in Figure~\ref{fig:conv1}: IRIS blocks from Paris's 11th~district converge much ``faster'' than those in the 8th~district, for instance.

\begin{figure}
\centering
\begin{minipage}[c]{.45\linewidth}
\begin{center}
\includegraphics[height=4.2cm]{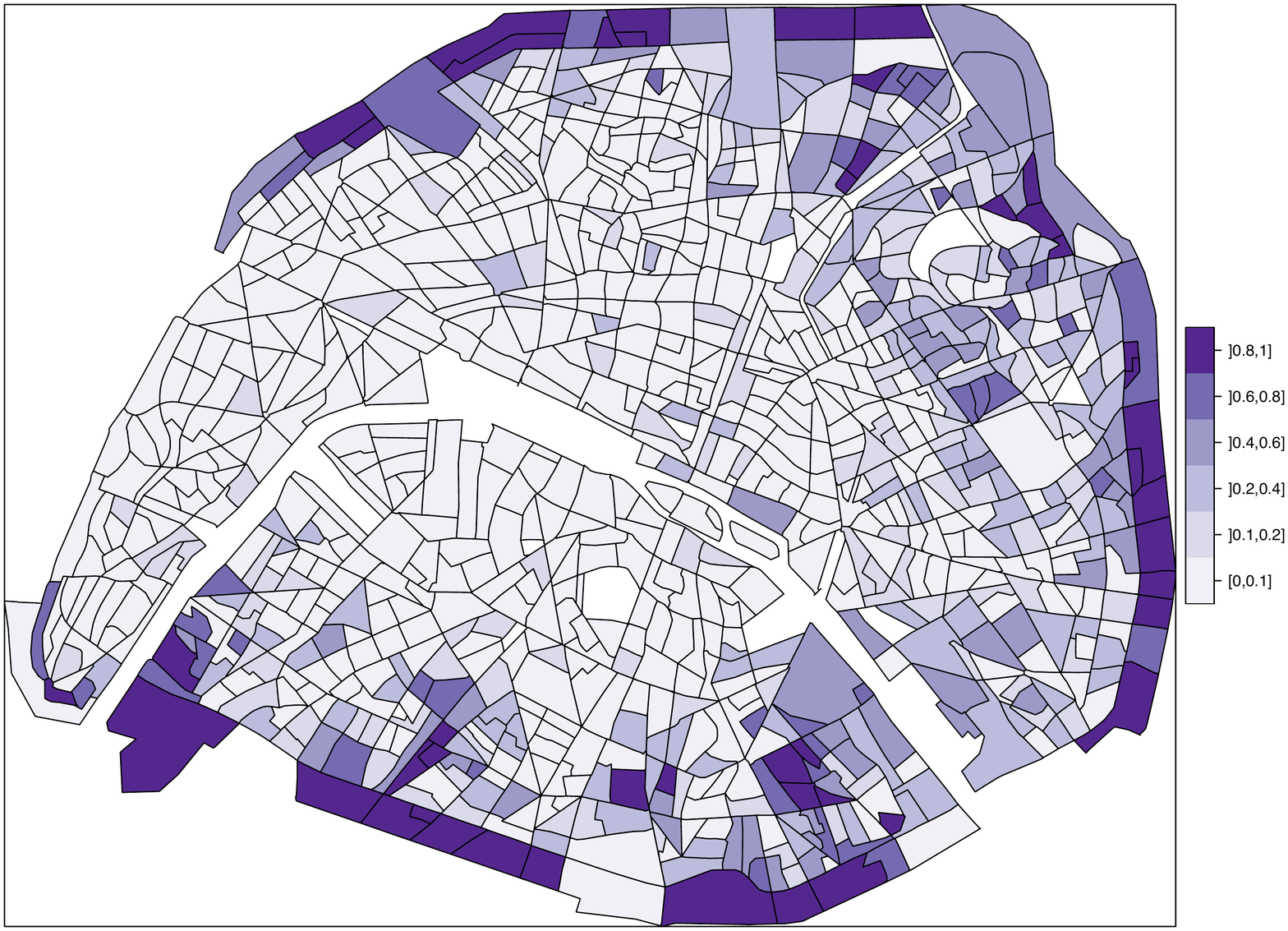}
\end{center}
   \end{minipage}
\begin{minipage}[c]{.45\linewidth}
\begin{center}
\includegraphics[height=4.2cm]{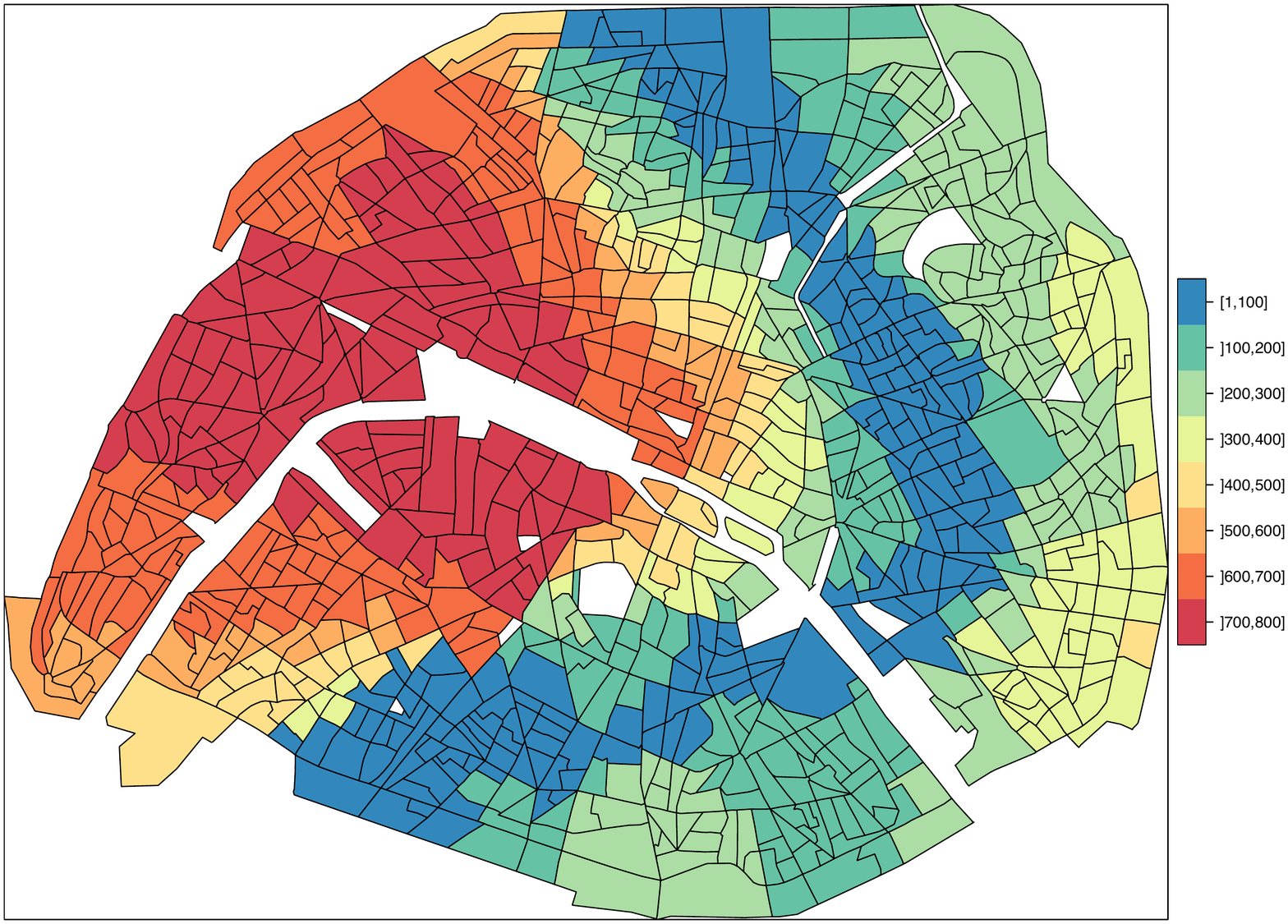}
\end{center}
   \end{minipage}
\caption{\label{fig:map}IRIS blocks coloured according to their radius of convergence to the global average social housing rate~(right). Radii are expressed in terms of the number of aggregation steps. For comparison, the map on the left-hand side shows IRIS areas coloured according to the local rate of social housing. (White areas correspond to parks, riverbanks and other IRIS where data is not available.) Observe that IRIS in the peripheral western part (W) and the peripheral eastern and north-eastern part (NE) of Paris present similar geographical positions and correspond to extreme values for the social housing rate (very high in the NE part, very low in the W part). Nonetheless, IRIS in the W part converge much slower to the city's mean, revealing a higher level of seclusion.}
\end{figure}

One may then classify IRIS blocks according to their radii of convergence to the city's mean, and represent them accordingly on a geographical map of Paris (Figure~\ref{fig:map}). First, note that boundary effects are clearly not predominant as peripheral western and north-eastern parts of the city, for instance, do not exhibit the same radii of convergence to the city's average. In fact, boundaries, by forcing the aggregation of IRIS blocks closer to the city centre and beyond (rather than neighbouring IRIS blocks just outside the city) tend to smooth patterns rather than exacerbate them.
A particularly interesting feature revealed by our method is that the western (W) part of the city is ``further away'' from the whole city than the north-eastern (NE) part: trajectories for IRIS in the W part need $2$ to $4$ times as many aggregation steps to converge to the city's average than IRIS in the NE part. Both parts correspond to extreme points for the variable in question, with a concentration of social housing in the NE part and a substantially lower than average rate in the W part (local social housing rates are also visible in Figure~\ref{fig:map}). Starting from these extremal points above and below the city's average, with symmetrical geographical position in the city, the W and NE parts could have had similar radii of convergence to the city's mean. However, this is clearly not the case, thus revealing a higher level of singularity, as far as social housing is concerned, in the W part of the city than in the NE part.

Conversely, one observes a quasi ring of IRIS blocks with relatively short scales of convergence to the city's mean. These are the historically socially mixed areas along the \textit{boulevards}, the former \textit{faubourgs} that used to be just outside the city's walls before these were transformed into \textit{boulevards}. Our method thus reveals a lasting imprint visible in terms of distance to the city's average for a variable, the social housing rate, that may be taken as a proxy to social diversity.

\subsubsection{Income distribution.}

We now consider a second example, and build trajectories for the income distribution in Paris.

As explained at the beginning of this section, when working with distributions available only through their quantiles, computing points for each group of blocks is a slightly more involved task. In this case, we chose to estimate the parameters of the best-fitting distribution (which happened to be log-normal) given a block's quantiles. Then, from this distribution, we simulated data corresponding to the number of households in the block. Thus we obtained a full ensemble of simulated households for every possible group of neighbouring blocks, from one block only to the whole city.
\medskip

\begin{figure}
\centering
\begin{minipage}[c]{.32\linewidth}
\begin{center}
\includegraphics[height=3cm,width=4cm]{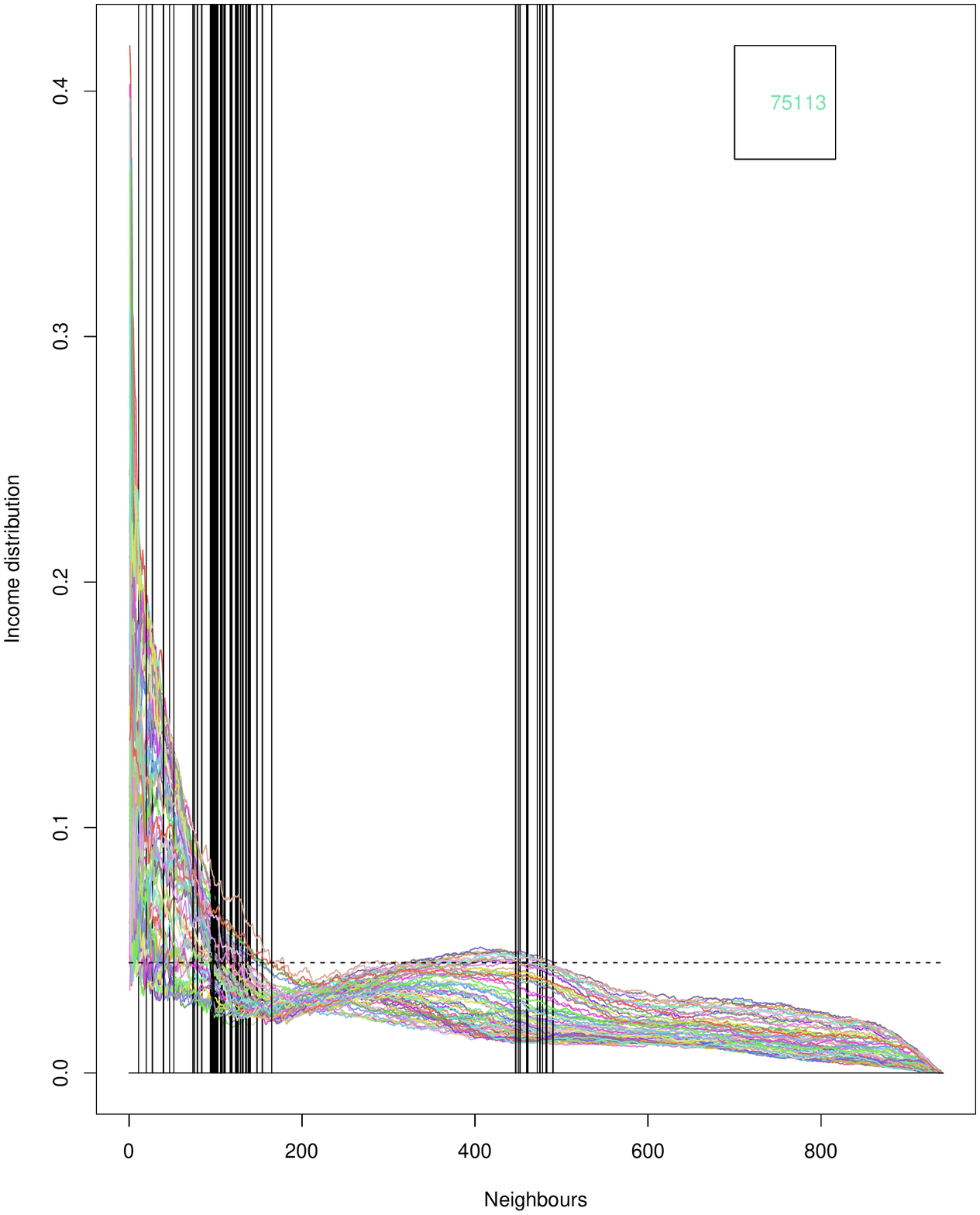}
\end{center}
   \end{minipage}
\begin{minipage}[c]{.32\linewidth}
\begin{center}
\includegraphics[height=3cm,width=4cm]{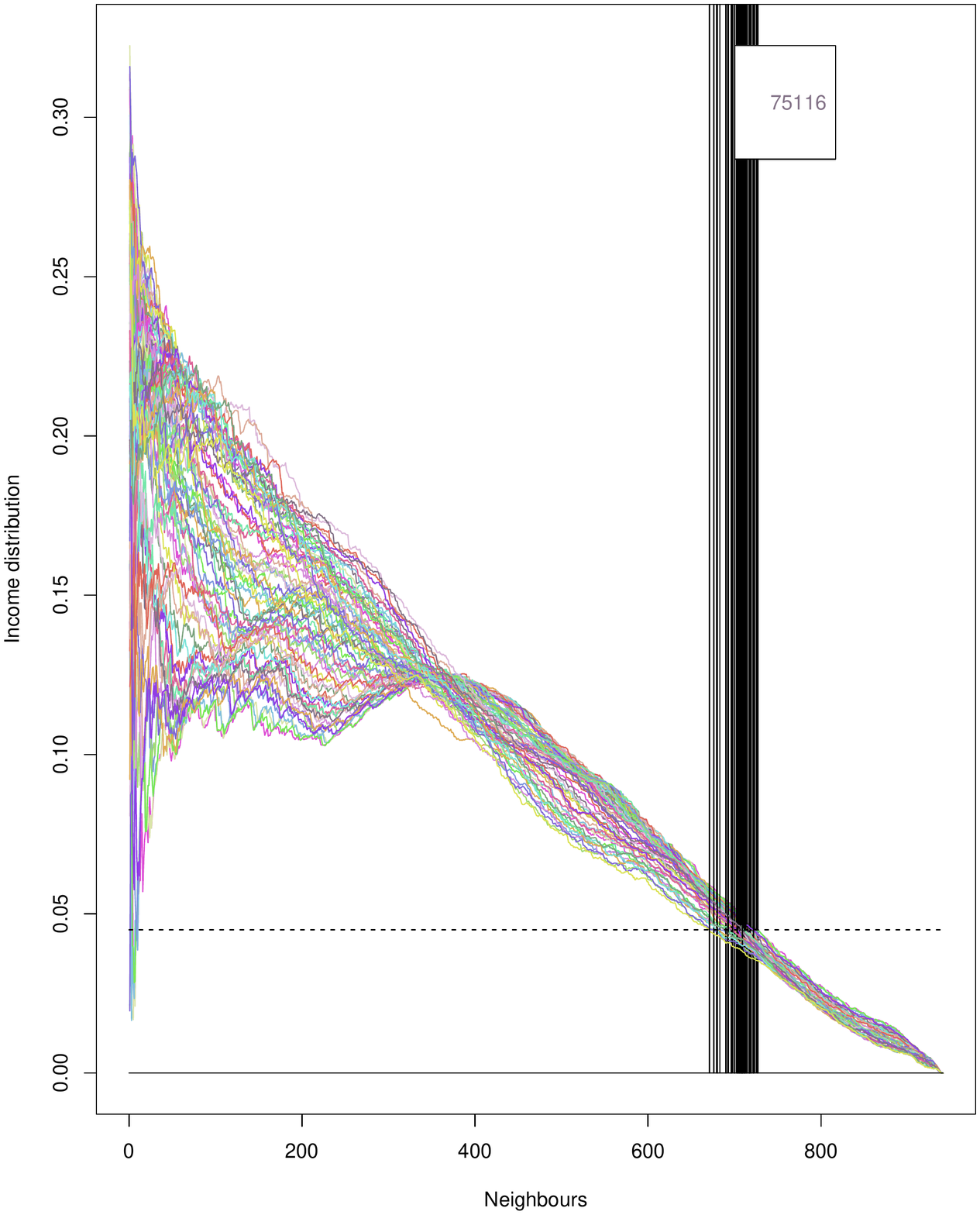}
\end{center}
   \end{minipage}
\begin{minipage}[c]{.32\linewidth}
\begin{center}
\includegraphics[height=3cm,width=4cm]{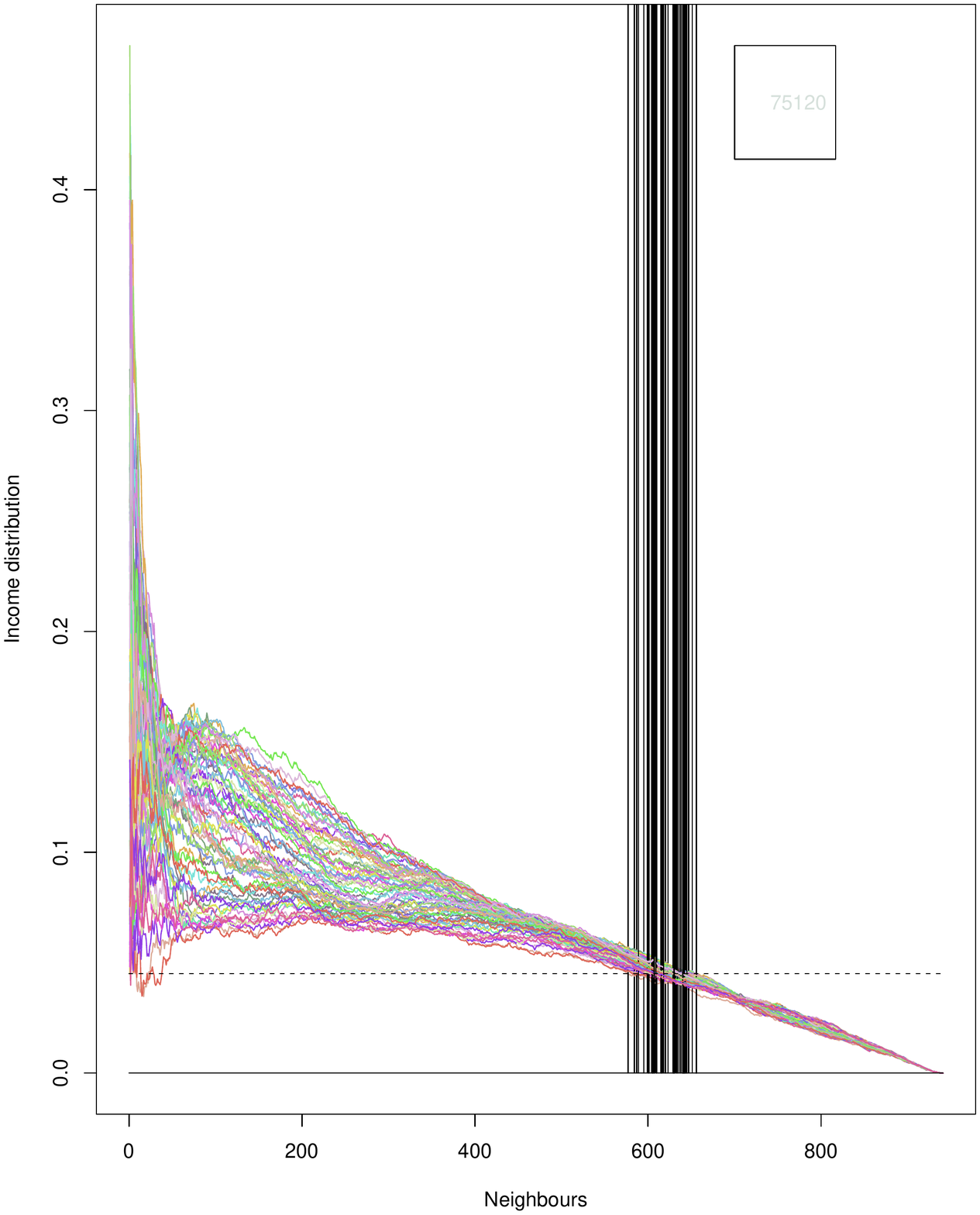}
\end{center}
   \end{minipage}
\caption{Trajectories for the income distribution, starting from each IRIS (census block) in Paris's 13th (left), 16th (middle) and 20th (right) districts. Ordinates correspond to the Kolmogorov-Smirnov distance (KS) between a group of blocks' income distribution and the whole city's income distribution. Solid vertical lines correspond to radii of convergence: the radius of convergence of a trajectory is defined as the point where the path last enters within $0.05$ of the city's average in terms of KS distance.}
\label{fig:ir1620}
\end{figure}

Let us look at trajectories for IRIS blocks in three Parisian districts of comparable sizes and population numbers: the 13th, 16th and 20th districts --~see Figure~\ref{fig:ir1620}. The first one corresponds to the south-western peripheral part of the city, the second one to the south-eastern peripheral part of the city, and the third one to the north-eastern peripheral part.

The 16th tends to be further away from the full city's picture, as it takes generally longer for its blocks to converge to the city's distribution than for blocks in the 13th or in the 20th districts. The 13th district exhibits an interesting behaviour, with some trajectories that first come close to the city's mean but then bounce up again further from it. This can be understood as follows: some IRIS blocks in the 13th district belong to relatively well-mixed neighbourhoods, so initially their trajectory approaches the city's average distribution. But the aggregation process around them leads to incorporate much less wealthy blocks on the border of the city, at a stage where wealthier blocks towards the center of the city and beyond have not yet been aggregated in large enough numbers to counterbalance the former. This sends trajectories away from the city's distribution again. A similar effect may (and does) take place for neighbourhoods relatively close to blocks that are much wealthier than the city as a whole (\textit{e.g.} in the north-western 17th district, not shown here).

\begin{figure}
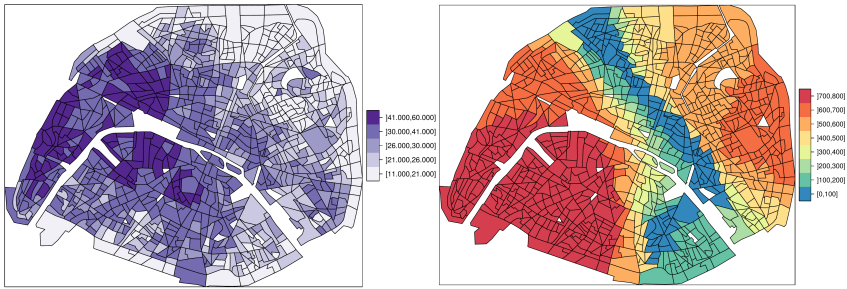

\centering
\begin{minipage}[c]{.45\linewidth}
\begin{center}
\includegraphics[height=3.8cm]{Revenus-Map.pdf}
\end{center}
   \end{minipage}
\begin{minipage}[c]{.45\linewidth}
\begin{center}
\includegraphics[height=3.8cm]{Map-Revenus.pdf}
\end{center}
   \end{minipage}
\caption{\label{fig:maprev}IRIS blocks coloured according to their radius of convergence to the city's income distribution~(right). For comparison, the map on the left-hand side shows IRIS areas coloured according to the median of the (local) income distribution. Distance between distributions is measured with the Kolmogorov-Smirnov distance. (White areas correspond to parks, riverbanks and other IRIS where data is not available.)}
\end{figure}
Let us now look at a map of Paris with IRIS blocks coloured according to their radius of convergence for the income distribution (Figure~\ref{fig:maprev}). The picture is here much more symmetric than for the social housing rate. Indeed, wealthier (south-western) and poorer (north-eastern) parts of the city both exhibit similar, slower convergence to the city's global distribution, compared to well-mixed areas such as the \textit{faubourgs}. However the wealthier, south-western part is again the last one to converge and constitutes a larger, more secluded area, including all of the 7th, 16th and 17th districts, and most of the 6th and (more surprisingly) the 14th district. Remarkably, the 8th district (around the \textit{Champs \'Elys\'ees}) does not stand out as a secluded area for the income distribution: this is because, geographically, the 8th is much closer to well-mixed neighbourhoods than, for instance, the 16th district.

\subsubsection{Metro and tramway station density.}

We come to a third type of data, for which one may define a local density. Recall the definition of
\begin{equation}
r_i(l)=\frac{m_i(l)/M}{p_i(l)/P},\label{eq:repind}
\end{equation}
where $M$ is the total number of metro and tramway stations in Paris, $m_i(l)$ the number of stations in a disk of radius $l$ centered on IRIS $i$, $p_i(l)$ the population living in the same disk, and $P$ the total population in Paris.\\
This quantifies whether the local density of metro stations per inhabitant is smaller ($r_i(l)<1$), larger ($r_i(l)>1$) or equal ($r_i(l)=1$) to the city's average density. In  other terms, as far as the number of metro and tramway stations is concerned, is the population living in the area under consideration deprived, well-served or neither?

Letting $l$ vary, we may then compute surface areas (equivalently, radii) for which, around a given point, one obtains a density per inhabitant similar to the average one in the whole city.

\begin{figure}
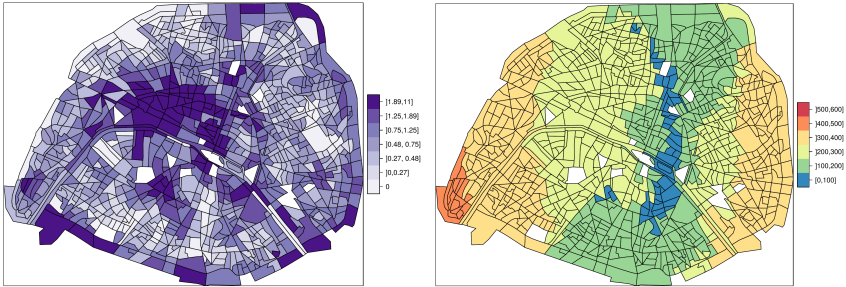

\centering
\begin{minipage}[c]{.45\linewidth}
\begin{center}
\includegraphics[height=3.8cm]{Transports-Map-ratio.pdf}
\end{center}
   \end{minipage}
\begin{minipage}[c]{.45\linewidth}
\begin{center}
\includegraphics[height=3.8cm]{Map-Transports.pdf}
\end{center}
   \end{minipage}
\caption{\label{fig:maptrans}IRIS areas coloured according to their radius of convergence to the city's average per-inhabitant density of metro and tramway stations~(right). For comparison, the map on the left-hand side shows IRIS areas coloured according to the local density of metro and tramway stations (within 500 meters of an IRIS's centroid). (White areas correspond to parks, riverbanks and other IRIS where data is not available.)}
\end{figure}

As can be seen in Figure~\ref{fig:maptrans}, our method provides a different picture from the one obtained by simply looking at local densities. The fact that the Parisian metropolitan network is mostly structured around a North-South axis is clearly visible in terms of convergence radii, whereas local densities almost exclusively show the higher concentration of metro stations in the city center. Note also that local densities are over-sensitive to the local presence of stations (\textit{eg} along tramway lines in peripheral IRIS blocks).

The quantity $r_i$ defined above (Eq.~\ref{eq:repind}) is similar to the ``representation'' defined in~\cite{BarthLouf} in order to examine patterns of residential segregation. Let us also follow in their footsteps by emphasizing, in the next section, the benefits of thinking in terms of what segregation is \textit{not}. Indeed, Louf and Barthelemy used their representation index to define a null model; we show here how our method of trajectories may be used to define similarly a reference model in terms of a uniform city.

\section{The uniform city as a null model: random walks to the mean}

The importance of null models has perhaps become more palpable in the last few decades, as new mathematical and statistical tools have come into play in a wider variety of fields, from urban geography to history and archeology (\cite{Filet17,Mills17}).\\
The null model against which segregation should be defined is the ``unsegregated'' city, as pointed out by~\cite{BarthLouf}. In the unsegregated city, distributions are uniform across space. So that, for instance, if the rate of social housing in the whole city is $\rho$, then any housing unit in the city, irrespective of its spatial position, has probability $\rho$ of being a social housing unit: there is not any spatial structure in the distribution of social housing units. This means that when considering a sequence of units defined solely according to some spatial rule (\textit{eg} aggregating sequentially nearest neighbours), no sign of this spatial rule will appear in the sequence: everything will be as if one were drawing units at random from a well-mixed urn.

Any deviation from such random behaviour will be a sign that an underlying structure leads to a biased shuffling of the units. Note that the underlying structure may be a spatial one, as in the viewpoint we are taking here, but it may also be of another nature depending on the type of data one is examining.

\subsection{A null model}
Let us examine here the somewhat ideal case when one is dealing with count data, i.e. data at the individual level. Suppose each individual belongs to one of two groups $A$ and $B$, \textit{eg} people belonging to one of two social groups, or housing units being of one of two types. Start from a given point in the city, and increase the population, one by one, from the individual located there to the whole population in the city. Write $X_i=1$ if the $i$-th individual is of type $A$, $X_i=0$ otherwise. Then setting $S_0=0$ and
\begin{equation}
S_n=\sum_{i=1}^{n}X_i
\end{equation}
produces a trajectory, that is: a sequence of points $(n,S_n)$, with $n$ varying from $0$ to the total number $N$ of individual units in the city. $S$ is simply the count data for group $A$.

In the case when the city is perfectly mixed, the trajectory thus produced will be a random walk $S$ that, given its current value $S_{n-1}$, either moves up with probability
\begin{equation}
\rho_n=\frac{\rho N-S_{n-1}}{N-n+1},
\label{eq:rho_n}
\end{equation}
or stays put with probability $1-\rho_n$, where $\rho$ represents the fraction of group $A$ in the total population~($N$). For the more mathematically inclined: such a random walk may also be viewed as a standard problem of drawing without replacement from an urn containing $N$ balls, $\rho N$ of which being of one type and the rest being of another type. $S_n$ would thus follow a well-known hypergeometric distribution -- see~\cite{feller1}.

Other sequences of interest that may be defined similarly are
\begin{itemize}
\item the sequence of averages: $$M_n=S_n/n;$$
\item the sequence of differences $D_n$, when one counts $+1$ for an individual of group $A$ and $-1$ otherwise: $$D_n=2S_n-n.$$
\end{itemize}
Both sequences, $M$ and $D$, are random walks that move up with probability $\rho_n$ or down with probability $1-\rho_n$, $\rho_n$ being as defined above in equation~(\ref{eq:rho_n}).

All three types of trajectories converge to well-identified final values: $S$ converges to $N\rho$, $M$ converges to $\rho$ and $D$ converges to $N(2\rho-1)$. From a probabilistic point of view, any of these three sequences is fully characterized by $\rho$ and $N$. Here they represent three facets of the same null model: a perfectly mixed city.

Now, if one has count data for a given variable in a city, one may build $N$ sequences (one for each starting point). One thus produces a sample of trajectories that would, if the city were well-mixed, exhibit the statistical properties of a set of $N$ random sequences obtained by drawing balls without replacement from a well-mixed urn. Any statistically significant deviation from such a set of random sequences signals the presence of spatial dissimilarities at multiple scales. We look at such deviations in the next subsection.

\subsection{Deviation from random sequences}
\label{sec:BT1}

\begin{figure}
\centering
\includegraphics[height=7cm]{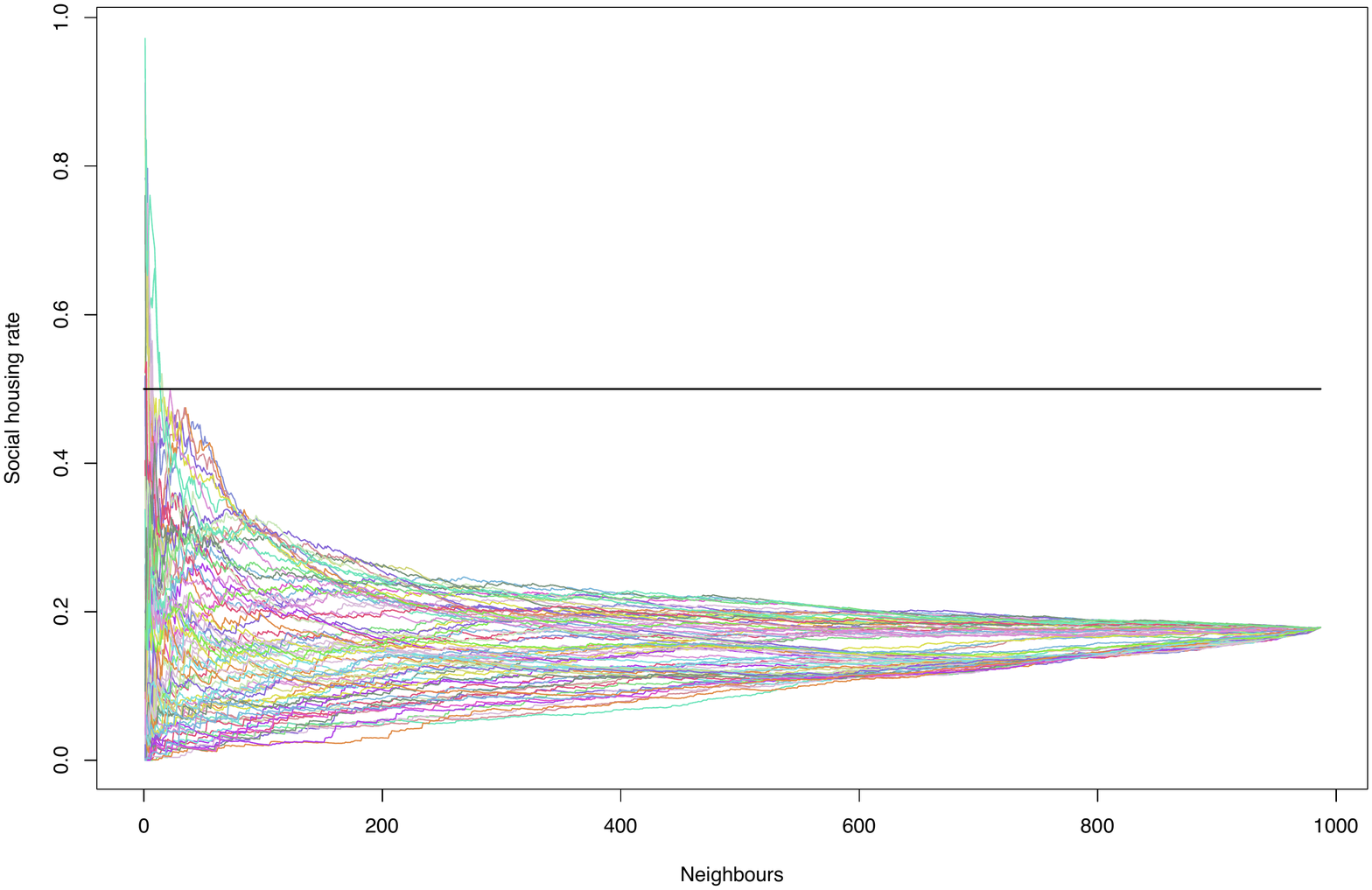}
\caption{Trajectories for the social housing rate, starting from some ($10\%$) of the 936 statistical units (so-called IRIS) in Paris. The solid line corresponds to a rate of $0.5$. According to the Ballot Theorem, were the city ``well-mixed'' in terms of social housing, just above $64\%$ of the trajectories would always remain below the $0.5$ mark. In Paris, almost $85\%$ of them do so.}
\label{fig:BT}
\end{figure}

Count data sequences are similar to sequences that one encounters when counting votes in an election. Imagine an election with two candidates, $A$ and $B$. Counting the votes cast for candidate~$A$ produces the $S$-sequence defined in the previous subsection, with $N\rho$ the final number of votes for~$A$. A famous result, or rather a famous set of results, linked to count data in elections, is the so-called Ballot Theorem:~\cite{whitworth86,bertrand87,feller1,takacs62,barton65,addario08}. In its simplest form, it gives the probability that $A$ be always in the lead, all through the counting process, given that $A$ wins the election with a final share $\rho$ of the votes. This is $2\rho-1$ (note that if $A$ wins the election, necessarily $\rho > 1/2$).

A very natural interpretation of the Ballot Theorem in terms of urban count data is the following. If I start from some point in the city and look at the first two housing units around me, and then the first three, and so on, what is the probability that I will always see a minority of social housing units, given that the fraction of social housing in the whole city is, say, $20\%$? The Ballot Theorem tells us this probability is $60\%$. So that, even if social housing is in a strong minority and if the city is perfectly mixed, when zooming out from a point chosen at random there is only a $60\%$ probability that one will never come across a bespoke neighbourhood where social housing is in the majority.

On count data, a first test for the presence of an underlying spatial structure therefore simply consists in examining significant deviation from the Ballot Theorem. In Paris for instance, the total fraction of social housing in the city is about $17.9$\%. Therefore $64.2$\% of the trajectories should always stay below the horizontal line $y=0.5$. In actuality, almost $85\%$ of them always stay below $0.5$ (Figure~\ref{fig:BT}).

Obviously, one should account for the fact that we only have a finite sample of sequences --~this is typically done through statistical hypothesis testing. In this specific context, we need a test to decide whether the observed deviation from the Ballot Theorem is significant or could be a simple random sampling effect. To this end, define the following sequence of Bernoulli random variables:
\begin{equation}
Y_i = \begin{cases} 1 &\mbox{if trajectory for the }i\mbox{-th IRIS is always below } 0.5 \\
0 & \mbox{otherwise. } \end{cases} \nonumber
\end{equation}
According to the Ballot Theorem,
\begin{equation}
\mbox{Prob}\left(Y_i = 1\right)=0.642, \quad \forall 1 \leq i \leq N. \nonumber
\end{equation}
If all $N$ trajectories were ``drawn'' independently from a well-mixed city, the sum $\sum_{i=1}^N Y_i$ would follow a binomial distribution with mean $0.642N$ and variance $0.23N$. Here a simple goodness-of-fit test shows that the Parisian trajectories for the social housing rate do not obey the Ballot Theorem, with a $p$-value of order $10^{-16}$\footnote{The minuteness of this $p$-value is due to the fact that the perfectly mixed city is an extreme, unrealistic case. It provides a scale that is too extreme, if one wishes to use it directly as a measure of segregation. However, it is an absolute scale, much in the same sense as the absolute temperature in physics. As such, it may be used reliably for comparisons.}.

Further, deviations from the null model of an unsegregated city may be characterized by a more complete statistical test. Indeed, with $N$ starting points in a city, one builds $N$ sequences, which should be compared to $N$ full drawings of a well-mixed urn. By full drawing, we mean drawing all balls, one by one, without replacement, until the urn is therefore left empty. In partial drawings, where one draws $n$ balls from an urn containing $N$ balls, standard statistical tests include the hypergeometric one, similar to Fisher's exact test~\cite{fisher25,fisher45,chvatal79}. In fact, one could use such a test at every single point of any of the trajectories. However, this would not be using the available information to the full. In future work, we will make use of a multiple test for the full set of trajectories' points $S_n^{(j)}$, with $1\leq n \leq N$ and $1\leq j \leq N$ --~see for instance~\cite{simes86}. We will also present ways of characterizing deviations from the null model for data types other than count data. This will lead to new ways of measuring segregation levels. A thorough, systematic investigation of these new indices, including rigorous comparisons with commonly used indices (see eg~\cite{masden88,fossett17}), will be carried out.

\section{Conclusion}
Aggregation of spatial data units has been so far generally considered as a difficulty to circumvene rather than an opportunity to analyze spatial dissimilarities. The new method introduced in this paper uses aggregation as a means to extract information on the relative singularity of each spatial unit within the city as a whole. Aggregation procedures may indeed reveal features that are not so easily seized from other perspectives. Another example is the recent use by~\cite{arcaute15} of aggregation to explore the behaviour (and more specifically the scaling laws) of various statistical variables across a phase space corresponding to almost all possible city boundaries' definitions in England and Wales. Aggregation is akin to an exploration of space from a given starting point, and it is this observation that forms the basis of the method we have presented here.

A question we have not enlarged upon in this paper is that of the definition of convergence. For instance, we chose a $\pm0.05$-interval for convergence to the city's average social housing rate. In fact, one may tune this threshold so as to study in greater detail areas that converge more slowly or more quickly. We will investigate this in future work. Other extensions that we will seek to develop include devising a visual representation of multigroup trajectories (other than a synthetic one such as a Kullback-Leibler distribution distance), and identifying relevant features on which to apply machine learning algorithms in order to obtain unsupervised clustering of trajectories. It will also be interesting to explore models able to reproduce stylized aspects of observed trajectories: hidden Markov models, multi-agent models, intermittent diffusion models.

We will also address a difficulty that arises in practice: one rarely has access to individual unit data. The data is often already aggregated at some basic statistical unit level. This means trajectories are observed only at certain points determined by the sequence of sizes of the statistical units instead of at every individual point. So that comparisons with Ballot Theorem results need to be made through what is called subordinated random walks --~which are precisely sequences observed at a number of different points rather than at every single step.

An interesting feature of our method is that there is no need to restrict oneself to the aggregation of nearest-neighbour units. One may also consider sequences defined from personal trajectories across the city (and thus aggregating units according to a personal ``distance'': units where one lives, works, goes out, etc). These personal trajectories transform the spatial network of blocks into personal networks reflecting personal paths and lives in the city. For some, the initial spatial network may turn into a small-world network, thanks to an ease of mobility across the different parts of the city. For others, personal trajectories may well be extremely concentrated around their local block. This is another aspect of segregation, that we will seek to explore, using data from surveys and geo-mobility (\textit{eg} mobile phone location data), and comparison with random walks across the spatial network.

\begin{acks}
The authors wish to thank Professor William A.V. Clark (UCLA) for his extremely helpful comments and remarks on a preliminary version of this article, and for many fruitful discussions.
\end{acks}

\begin{funding}
This research received no specific grant from any funding agency in the public, commercial, or not-for-profit sectors. 
\end{funding}

\bibliographystyle{SageH}
\bibliography{SchellingBib.bib} 

\small
\begin{biogs}
\textbf{Dr Julien Randon-Furling} is a Senior Lecturer in Mathematical Sciences at Universit\'e Paris 1 Panth\'eon Sorbonne.

\noindent \textbf{Dr Madalina Olteanu} is a Senior Lecturer in Statistics at Universit\'e Paris 1 Panth\'eon Sorbonne, currently on sabbatical leave at INRA, Universit\'e Paris-Saclay.

\noindent \textbf{Antoine Lucquiaud} is a doctoral researcher at the centre for Statistics, Analysis, and Multidisciplinary Modelling, Universit\'e Paris 1 Panth\'eon Sorbonne.
\end{biogs}

\end{document}